\documentclass{aa}
\usepackage{graphicx}
\usepackage{txfonts}
\def \inte {INTEGRAL}

\def \src {IGR~J11215--5952}

\def \hcm {\hbox {\ifmmode $ atom cm$^{-2}\else atom cm$^{-2}$\fi}}

\begin{document}
   \title{IGR J11215--5952: a hard X--ray transient displaying recurrent outbursts\thanks{Based on
   observations with INTEGRAL, an ESA project with instruments and the science data
   centre funded by ESA member states (especially the PI countries: Denmark, France,
   Germany, Italy, Switzerland, Spain), Czech Republic and Poland, and with the
    participation of Russia and the USA.} }


   \author{L. Sidoli\inst{1}
                \and
      A. Paizis\inst{1}
          \and
          S. Mereghetti\inst{1}
}

   \offprints{L. Sidoli, sidoli@iasf-milano.inaf.it}

   \institute{INAF, Istituto di Astrofisica Spaziale e Fisica Cosmica  Milano,
   via E.Bassini 15, I-20133 Milano, Italy
             }

   \date{Received January 31, 2006; accepted March 2, 2006}


\abstract
  {The hard X--ray source \src\ has been  discovered with INTEGRAL
during a short outburst in 2005 and proposed as a new member of
the class of supergiant fast X-ray transients.}
   {We analysed \inte\ public observations of the source field
in order to search for  previous outbursts from this transient, 
not reported in literature. 
}
  {Our results are based on a systematic re-analysis of \inte\
archival observations, using the latest analysis software and
instrument calibrations. }
   {We report the discovery of two previously unnoticed 
outbursts,  spaced by intervals of $\sim$330 days, 
that occurred in July 2003 and May
2004.
The 5--100~keV spectrum of \src\ is well described by a cut-off
power law, with a photon index of $\sim$0.5, and a cut-off energy
$\sim$15--20~keV, typical of High Mass X--ray Binaries hosting a
neutron star. A  5--100~keV luminosity of
$\sim$3$\times$10$^{36}$~erg~s$^{-1}$ has been  derived (assuming
6.2~kpc, the distance of the likely optical counterpart).}
  {The 5--100~keV spectral properties, the recurrent nature
of the outbursts,
together with the reduced error region containing the  blue
supergiant star HD 306414,  support the hypothesis that \src\ is a member of
the class of the Supergiant Fast X--ray Transients. 
}


   \maketitle
%

\section{Introduction}

More than one hundred hard X--ray (E$>15$~keV) sources have been
discovered with the INTEGRAL satellite since its launch in 2002
(see e.g. Bird et al. 2005). About one third of these new sources
are associated with High Mass X--ray Binaries (HMXRBs), either
thanks to their secure identification at optical/infrared
wavelengths with blue supergiants or Be stars, or based on their
X--ray properties typical of HMXRBs, like e.g. periodic pulsations
or hard X--ray/gamma--ray spectra.

The discovery of several transient sources with Be star companions
was not unexpected, since these sources are thought to represent
the most abundant class of HMXRBs in the Galaxy. Every new X-ray
satellite mission increased the sample of these objects, by
detecting outbursts from new sources. More interestingly, INTEGRAL
discovered two other kinds of HMXRBs with supergiant companions
that escaped detection with previous high-energy satellites:
highly absorbed persistent sources, often showing periodic
pulsations (e.g., Lutovinov et al. 2005), and recurrent transients
characterized by short outbursts lasting only a few hours (e.g.
Negueruela et al. 2005a,  Sguera et al. 2005).

The transient hard X--ray source \src\ was discovered with the
INTEGRAL satellite in  April 2005 (Lubinski et al. 2005)
and tentatively associated with the supergiant star HD~306414
(Negueruela et al. 2005b). In the course of a systematic
re-analysis of all the observations from the INTEGRAL public data
archive we discovered two previous outbursts of \src\ which
demonstrate the recurrent nature of this transient and hint to a
possible periodicity of about 330 days for its outbursts.


\begin{table*}
\begin{center}
\caption{Summary of the Science Windows with detections 
of  \src\ }
\label{tab:det}
\centering
\begin{tabular}{lllll}     
\hline
\hline                     
Obs ID         & IJD  & Detection significance$^{(a)}$  & ISGRI rate$^{(a)}$ & Off-axis angle  \\
               & (=MJD--51544)     &   $\sigma$ &  (s$^{-1}$)  &    (degrees)      \\
\hline
First outburst   (3-4 July 2003)    &    &   &   &  \\
\hline
008800020010   &   1279.76 &     6.8   &    3.6  &   4.4 \\
008800040010   &   1279.81 &     9.3   &    3.8  &   4.8 \\
008800060010  &    1279.87  &     5.4   &    3.0  &   6.5  \\
008800080010  &    1279.91 &     6.4   &    3.3  &   2.6  \\
008800090010  &    1279.94 &     6.0   &    3.0  &   3.4  \\
008800120010  &    1280.01 &     8.5   &    4.6  &   5.1  \\
008800130010  &    1280.03 &     9.2   &    4.9  &   3.5  \\
008800140010  &    1280.05 &     7.3  &   3.9   &   2.6  \\
008800150010  &    1280.07 &     9.1   &   4.6   &   4.5  \\
008800170010  &    1280.12 &     8.2  &    4.7  &    6.8  \\
008800180010  &    1280.14 &     8.2   &   4.5   &   4.8  \\

\hline
Second outburst  (26-27 May 2004)    &    &   &   &  \\
\hline
019700600010   &   1608.09 &     8.7  &    3.1 &     1.4 \\
019700610010   &   1608.13 &     8.6  &    3.2 &     2.4 \\
019700640010   &   1608.27 &     5.8  &    2.9 &     7.6 \\
019800050010   &   1609.00 &     7.0  &    2.9 &     6.8 \\
019800070010   &   1609.08 &     7.2  &    3.9 &     9.5 \\
019800320010   &   1610.12 &     5.7  &    2.2 &      2.5 \\
\hline
\end{tabular}
\end{center}
\begin{small}
$^{(a)}$ in the 17-40 keV energy range \\
\end{small}
\end{table*}

\section{Observations and Results}

The ESA INTEGRAL gamma-ray observatory, launched in October 2002,
carries three co-aligned coded mask telescopes: the imager IBIS
(Ubertini et al. 2003), which allows high angular resolution
imaging over a large field of view (29$^{\circ}\times29^{\circ}$)
in the energy range 15\,keV--10\,MeV, the spectrometer SPI
(Vedrenne et al. 2003; 20 keV--8\,MeV) and the X-ray monitor JEM-X
(Lund et al. 2003; 3--35\,keV).  IBIS is composed of a low-energy
CdTe detector (ISGRI; Lebrun et al. 2003), sensitive in the energy
range from 15\,keV to 1\,MeV, and a CsI detector (PICsIT; Labanti
et al. 2003), designed for optimal performance at 511 keV, and
sensitive in the 175\,keV--10\,MeV energy range.

The sky region of \src\ was repeatedly observed by INTEGRAL. We
analyzed all the public ISGRI observations pointed within
15$^{\circ}$ of the source, i.e. 850 individual pointings (Science
Windows) performed between December 2002 and August 2004,
yielding  a total exposure time of about 1.8~Ms. We processed all
the data using version 5.1 of the OSA  \inte\ analysis software,
with the corresponding response matrices.

\src\ was  detected  with a significance larger than 5$\sigma$ in
the 17--40 keV range in  17 Science Windows. These detections
correspond to two 
outbursts occurring on 3-4 July 2003 and
26-27 May 2004 (see Table~\ref{tab:det}). 
Due to the sparse sampling of the
observations, we  cannot determine precisely the duration
of the outbursts. The lower limits on their duration are
 $\sim$9 hours and about two days for the 2003
and 2004 outbursts, respectively.
We can also assess that the duration of both 
outbursts did not exceed $\sim$7~days.

\begin{figure}
  \centering
   \includegraphics[angle=0,width=8.7cm]{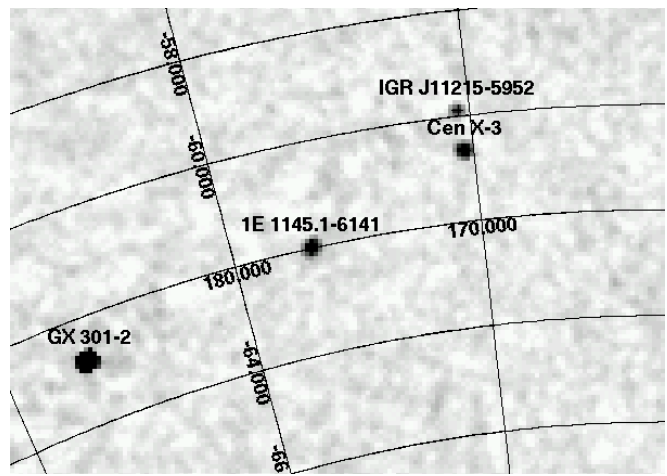}
      \caption{Section (about 10$^{\circ}\times8^{\circ}$) of the ISGRI mosaic
      of the two outbursts
       in the energy range 17--40~keV, with a net exposure time of 42.7~ks. Equatorial coordinates
(J2000) are displayed. 
              }
         \label{fig:pos}
   \end{figure}

\begin{figure}
\hspace{-1.cm}
   \includegraphics[angle=0,width=11.cm]{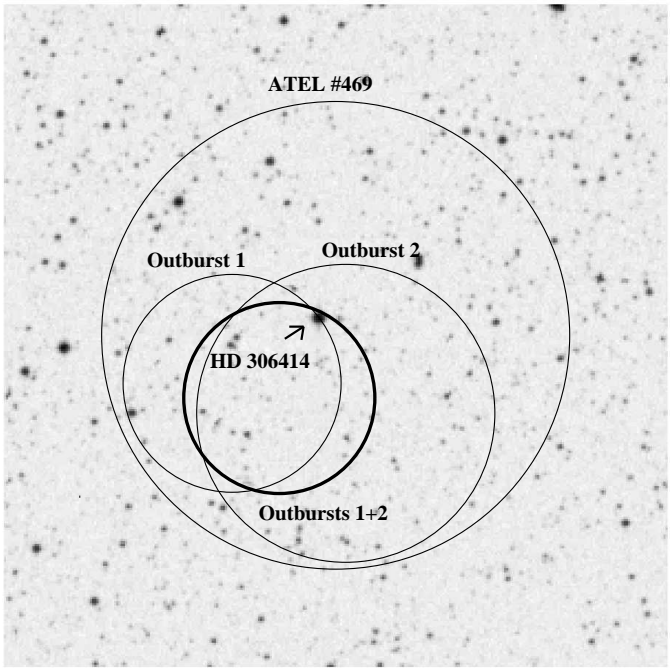}
      \caption{ J--band image of the \src\ field (data taken from UK Schmidt telescope,
and provided by the online Digitized Sky Survey, available at {\em http://heasarc.gsfc.nasa.gov/}).
North to the top, East to the left. 
The large solid 
circle marks the ISGRI error circle (3$'$ radius) from the 2005 outburst (Lubinski et al. 2005, ATel 496);
the two thin smaller circles indicate the  error regions from the analysis of the 
outbursts occurred in 2003 and 2004. 
The smaller thick solid circle indicates the error region derived in
the mosaic  of   the two outburst in 2003 and 2004.
The arrow indicates HD~306414, the likely optical counterpart (Negueruela et al. 2005).
              }
         \label{fig:poserr}
   \end{figure}

Combining the Science Windows of the two outbursts in a single
mosaic we obtained the image  shown in Fig.~\ref{fig:pos}.  \src\
is clearly visible  at $\sim45'$ from the bright X-ray pulsar 
Cen X-3. 
Its coordinates (J2000) are: R.A.= 11$^h$~21$^m$~50.8$^s$,
Dec.= --59$^{\circ}$~52$'$~48.3$''$, with a statistical uncertainty of 1.2$'$.
This refined position is consistent with what measured during the April
2005 discovery outburst (Lubinski et al. 2005) and our smaller
error region still contains the proposed optical counterpart 
HD~306414 (Masetti et al. 2006).
The different error regions are compared in Fig.~\ref{fig:poserr}.

Based on  our refined source position, we extracted the ISGRI and JEM-X spectra.
The latter were available only for a subset of the Science
Windows, due to the smaller field of view of the JEM-X instrument
(radius 3.5$^\circ$).

\begin{figure*}[ht!]
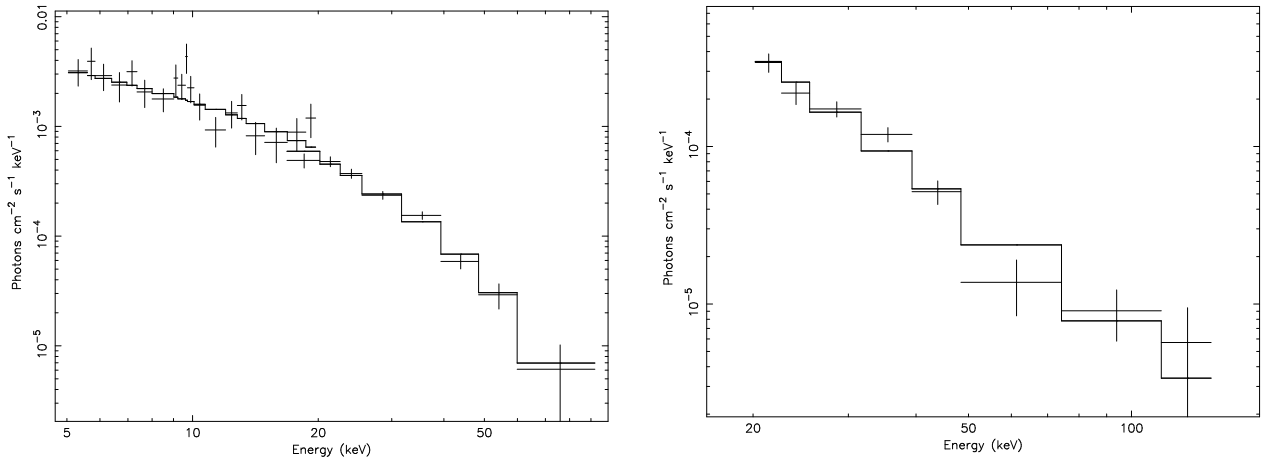

\hbox{\hspace{.3cm}
\includegraphics[height=8cm,angle=-90]{4940fig3a.ps}
\hspace{.5cm}
\includegraphics[height=8cm,angle=-90]{4940fig3b.ps}}
\caption[]{{\em Left:}  \src\  photon spectrum  from the 2003
outburst (JEM-X and ISGRI, 5--100 keV) fit with a cutoff power
law. {\em Right:}  \src\  photon spectrum  (only ISGRI data,
20--100~keV) from the 2004 outburst, fit with a single power law
(see the text for the best fit parameters). }
 \label{fig:spec}
\end{figure*}

\begin{figure}[ht!]
   \centering
   \includegraphics[angle=-90,width=8.0cm]{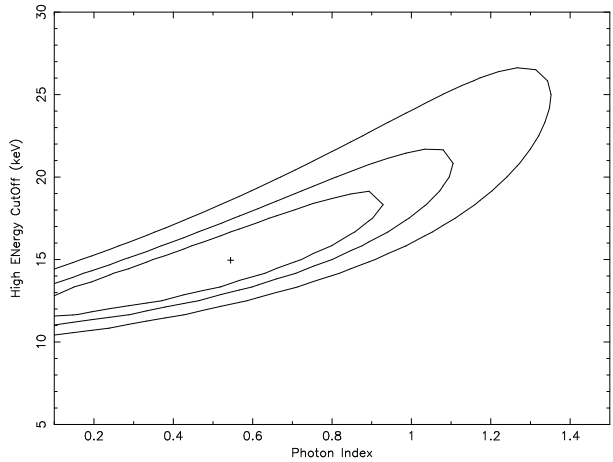}
      \caption{68\%, 90\% and 99\% confidence
level contours for the photon index and the cutoff energy from the
spectral analysis of the emission during the 2003 outburst.
              }
         \label{fig:cont}
   \end{figure}

The source spectra extracted at the peak of the two different
outbursts, are plotted in Fig.~\ref{fig:spec}. 
The 5--100~keV spectrum of July 2003 (Fig.~\ref{fig:spec}, left panel) 
could not be fitted with a
single power law, while a
cutoff power law gave an acceptable result ($\chi^2$=47.5 for 50
degrees of freedom, dof), with a photon index of 0.5$^{+0.4}
_{-0.6}$ and cutoff energy of 15$^{+5} _{-4}$~keV. In
Fig.~\ref{fig:cont} we show the contour plot for the photon index
and the cutoff energy in this model. The best fit parameters
correspond to fluxes of
6.2$\times$10$^{-10}$~erg~cm$^{-2}$~s$^{-1}$ (5--100 keV) and
2.8$\times$10$^{-10}$~erg~cm$^{-2}$~s$^{-1}$ (20--60 keV). An
equally good fit could be obtained with a thermal bremsstrahlung
model with  temperature of 26$^{+5} _{-4}$~keV.

The ISGRI spectrum observed during the second outburst (May 2004) 
is shown in
the right hand panel of Fig.~\ref{fig:spec}. 
The  JEM-X spectrum could not be used because the
instrument was switched off during one of the Science Windows in
which the source was nearly on-axis, while in the other Science
Windows  the low statistics hampered a meaningful spectral
extraction.
Due to the lack
of JEM-X spectrum, ISGRI spectrum can be fitted with
a single power law ($\chi^2$=12.9 for 8 dof) with a photon index
of 2.6$^{+1.8} _{-0.6}$. The 20--100~keV flux is
2.5$\times$10$^{-10}$~erg~cm$^{-2}$~s$^{-1}$ and the 20--60~keV
flux is 2.1$\times$10$^{-10}$~erg~cm$^{-2}$~s$^{-1}$.

\section{Discussion and Conclusions}

Using all the available public INTEGRAL data, we have discovered
two  outbursts and obtained a refined position for the
transient hard X-ray  source \src. The new error circle is still
consistent with the proposed optical counterpart HD~306414, which
has been recently studied by Masetti et al. (2006). These authors
found evidence for an H$_{\alpha}$ emission line, confirmed  the
spectral classification as a  B1~Ia-type star and estimated a
distance of d$\sim$6.2~kpc, placing HD 306414 in the Carina spiral arm. 
Unfortunately, the JEM-X data do not extend to
sufficiently low energy to constrain the interstellar absorption
and thus give some indication on the distance of the X-ray source.

For the distance of 6.2 kpc, the peak fluxes of the two outbursts
reported here correspond to a  luminosity of
$\sim$3$\times$10$^{36}$~erg~s$^{-1}$ (5--100~keV). This
luminosity, as well as the spectral shape derived with INTEGRAL
are typical of  High Mass X--ray Binaries containing a neutron
star.

The transient nature of the source, the spectral properties observed
during the two outbursts, together with the  blue
supergiant companion, suggest that \src\ is likely a member
of the class of Supergiant Fast X--ray Transients (SFXTs; e.g.,
Smith et al. 2006; Negueruela et al. 2005a). 
This is a recently recognized class of X--ray
binaries with a supergiant companion, similar for what concerns
their spectral properties to  the persistent X--ray binary
pulsars, but characterized by the emission of a significant
luminosity only during short X--ray outburst. This behavior is
quite surprising since neutron stars accreting from the winds of
supergiant companions were, until recently, seen as relatively
steady sources. 
The two newly discovered outbursts from \src\ are
somehow longer than typical outbursts observed to date from other
SFXTs (which are shorter than 3~hours, Sguera et al. 2005), but
are not unusual (see, e.g., the outbursts  observed 
from the SFXT XTE~J1739-302, Smith et al. 2006).

The three outbursts observed to date from \mbox{\src}, the two reported
here and the one that led to the source discovery (22 April 2005;
data not yet in the public archive), are consistent with a
recurrence time of about 330 days\footnote{We cannot exclude that
the true periodicity is  half of this period, because of the lack
of \inte\ continuous coverage of the source position}. This
possible periodicity, if confirmed, is worth noting, since in no
other source belonging to the class of SFXTs a periodic behavior
has been observed. Indeed, only the recurrent (but not periodic)
nature of the outbursts from few members of the class has been
observed before (Sguera et al. 2005). The possible $\sim$330~days
periodicity could represent the orbital period of the binary
system, although such a long period is more typical of Be/X--ray
binaries than of Supergiant HMXRB, which, in general, have orbital
periods shorter  than $\sim$20~days. Negueruela et al. (2005a)
suggested that SFXTs have wider orbits than ``normal'' supergiant
persistent HMXRBs (Vela X--1-like systems) and that the compact
source accretes from a less dense environment, in order to explain
the very low emission level during quiescence in SFXTs
($\sim$10$^{32}$--10$^{33}$~erg~s$^{-1}$). Our possible finding of
a long periodic recurrence time of the outbursts could confirm
this hypothesis.

\begin{acknowledgements}

This work has been partially supported by the Italian Space Agency
and by the MIUR under grant PRIN 2004-023189.

\end{acknowledgements}


\begin{thebibliography}{}

\bibitem[]{}Bird, A.J., Barlow, E.J., Bassani, L., et al., 2006, ApJ, 636, 765

\bibitem[]{}Kuulkers, E., 2005, in ``Interacting Binaries: Accretion, Evolution and Outcomes",
Eds. L.A. Antonelli, et al., Proc. of the Interacting Binaries Meeting of Cefalu,
Italy, July 2004, AIP, in press

\bibitem[]{}Labanti, C., Di Cocco, G., Ferro, G., et al. 2003, A\&A, 411, L149

\bibitem[]{}Lebrun, F., Leray, J.P., Lavocat, P., et al. 2003, A\&A, 411, L141

\bibitem[]{}Lubinski, P., Gadolle Bel, M., von Kienlin, A., et al., 2005, ATel 469

\bibitem[]{}Lund, N., Budtz-J{\o}rgensen, C., Westergaard, et al. 2003, A\&A, 411, L231

\bibitem[]{}Lutovinov, A., Rodriguez, J., Revnivtsev, M., et al., 2005, A\&A, 433, L41

\bibitem[]{}Masetti, N., Pretorius, M.L., Palazzi, E., et al., 2006, A\&A in press (astro-ph/0512399)

\bibitem[]{}Negueruela, I., Smith, D.M., Reig, P., et al., 2005a, astro-ph/0511088

\bibitem[]{}Negueruela, I., Smith, D.M., Chaty, S., 2005b, ATel 470

\bibitem[]{}Sguera, V., Barlow, E.J., Bird, A.J., et al., 2005, A\&A, 444, 221

\bibitem[]{}Smith, D.M., Heindl, W.A., Markwardt, C.B., et al., 2006, ApJ, 638, 974

\bibitem[]{}Ubertini, P., Lebrun, F., Di Cocco, G., et al. 2003, A\&A, 411, L131

\bibitem[]{}Vedrenne, G., Roques, J.-P., Sch{\" o}nfelder, V., et al. 2003, A\&A, 411, L63


\end{thebibliography}
\end{document}